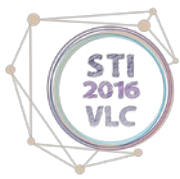

# Can we use altmetrics at the institutional level? A case study analysing the coverage by research areas of four Spanish universities[1]

Daniel Torres-Salinas[*], Nicolas Robinson-Garcia[**] and Evaristo Jiménez-Contreras[***]

[*] torressalinas@gmail.com
Universidad de Navarra and Universidad de Granada (EC3metrics and Medialab UGR), Granada (Spain)

[**] elrobin@ingenio.upv.es
INGENIO (CSIC-UPV), Universitat Politècnica de València, Valencia (Spain)

[***] evaristo@ugr.es
Universidad de Granada, Dpto. Información y Comunicación (Grupo EC3 and EC3metrics), Granada (Spain)

## INTRODUCTION

Social media based indicators or *altmetrics* have been under scrutiny for the last seven years. Their promise as alternative metrics for measuring scholarly impact (Priem, Piwowar, & Hemminger, 2012) is still far from becoming a reality (Torres-Salinas, Cabezas-Clavijo, & Jiménez-Contreras, 2013). Issues with regard to their meaning (Sud & Thelwall, 2014), potential use as alternative or complements to citation indicators (Bornmann, 2014; Costas, Zahedi, & Wouters, 2015a), data collection inconsistencies (Robinson-García, Torres-Salinas, Zahedi, & Costas, 2014) or diversity of sources (Thelwall, Haustein, Larivière, & Sugimoto, 2013) are still under development. In this regard, three lines of work can be observed in regard to altmetric studies: 1) the relation between citations and altmetric indicators (Costas et al., 2015a), 2) their meaning as impact indicators (i.e., Bornmann, 2014; Haustein, Bowman, & Costas, 2015), and, 3) coverage and diversity of social media sources (i.e., Haustein, 2016).

Altmetric indicators have been intimately related since their conception with commercial interests (Bornmann, 2014) and many data providers offering social media metrics have been developed. Here Altmetric.com has positioned itself as the most spread and used data sources for altmetric studies. Altmetric.com has the advantage of providing detailed data from a variety of social media platforms with regard to mentions, readership, etc. of scientific papers. It also offers an aggregated indicator or 'altmetric' score which is based on a weighted sum of values based on the presence of a given article in different social media. Contrarily to its competitors, it has the advantage of maintaining the 'history' of each altmetric indicator to an article, thus ending with a classical limitation of this type of indicators: their volatility (Costas et al., 2015a).

Up to now, most studies have focused on the understanding of the nature and relation of altmetric indicators with citation data. Few papers have analysed research profiles based on altmetric data. Most of these have related to researcher profiles and the expansion of these tools among researchers. For instance, (Haustein et al., 2014) surveyed participants of the STI2012 Conference to learn the spread on the use of these tools by researchers. With a similar aim but different approach, (Torres-Salinas & Milanés-Guisado, 2014) analysed the presence in social media of the most prolific authors in the Profesional de la Información journal. Finally, (Martín-Martín, Orduna-Malea, & Delgado López-Cózar, 2016) created a portal including altmetric and citation scores for profiling researchers according to the source from where data is retrieved. However, no study has been found profiling larger units of research such as institutions, especially universities. Despite how common it is to find traditional bibliometric studies analysing institutional research performance, currently there is no study focused on the utility, consistency and, especially the thematic coverage of altmetrics at the institutional level.

---

[1] Nicolas Robinson-Garcia is currently supported by a Juan de la Cierva-Formación Fellowship from the Spanish Ministry of Economy and Competitiveness.

This paper aims at exploring the coverage of the Altmetric.com database and its potential use in order to show universities' research profiles in relationship with other databases. Specifically, our objectives are the following:

1. Analyse the coverage of Altmetric.com at the institutional level and verify its validity as a data source for obtaining alternative metrics derived from the research activity of universities in comparison with those from the Web of Science. For this, we will work with a small sample of four Spanish universities with different characteristics.

2. Analyse coverage differences when obtaining bibliometric profiles from Altmetric.com and Web of Science. In some studies a higher coverage of the Social Sciences and Humanities has been reported, suggesting the potential of altmetric indicators in these areas (Costas, Zahedi, & Wouters, 2015b).

## MATERIAL AND METHODS

In this paper we will analyse a sample of four different Spanish universities. These universities present different sizes and research publication specialization as well as being located in different regions of the country. Table 1 shows a description of the characteristics of each university, the main field in which they are specialised, foundation year, and total number of publications produced in 2014 according to the Web of Science database.

Table 1. Description of the four universities analysed in this study

| *University* | *Size Staff Full Time* | *Type and main field* | *Foundation Year* | *Region & City* | *Nº Web of Science Docs 2014*|
|---|---|---|---|---|---|
| University of Granada | 2399 | **Multidisciplinary** | 1531 | Andalucía Granada | 2387 |
| University Pompeu Fabra | 288 | **Specialized** Medicine/Biology | 1990 | Cataluña Barcelona | 1029 |
| Polytechnic University Valencia | 1847 | **Specialized** Engineering | 1971 | Comunidad Valenciana Valencia | 1753 |
| University Carlos III | 539 | **Specialized** Social Sciences | 1989 | Comunidad Madrid Madrid | 810 |

All publications for the 2014 year indexed in the Web of Science citation indexes (SCI, SSCI and H&ACI) as articles, reviews, notes or letters for our sample of institutions were retrieved in November 2015. A total of 5922 records was retrieved. Records from the Web of Science provide among other information the DOI number of each publication, although this is not necessarily provided for all documents. The DOI number is very important, as it will allow us to query the Altmetric.com API for each document (Robinson-García et al., 2014). A total of 5547 papers from our data set included DOI (93% of the total share). All universities showed a similar share of documents with DOI number except for the University of Granada, where the percentage went down to 91%. Finally, we retrieved all altmetric data available at the moment for our set of papers with DOI.

Publications from the Web of Science have been assigned to four broad research areas: "Engineering & Technology", "Humanities & Arts", "Science" and "Social Sciences". These areas have been constructed by aggregating Web of Science Subject Categories. Then, the share of records with a score assigned by Altmetric.com (as defined above) has been computed. The range of such score for our data set was of 928.

The data set employed for this study is publicly available for reproducibility purposes at https://dx.doi.org/10.6084/m9.figshare.3120946.v3. This data set includes all records retrieved from the Web of Science as well as social media metrics and score retrieved from Altmetric.com along with their assigned research area.

## RESULTS

Table 1 shows the distribution of publications by research area in order to learn the research profile of each university. The University of Granada is the largest university of the four analysed with up to 2387 papers published in 2014, followed by Polytechnic University of Valencia, Pompeu Fabra University and Carlos III University. Regarding their research profile, we first must note that Science in all cases is the area with the highest share of output. The only exception is Carlos III University, where Engineering & Technology represent 51% of the total share followed by 45% of Science. Engineering has a large presence also in the Polytechnic University of Valencia. Social Sciences reach their highest presence in Carlos III University (20%), followed by Pompeu Fabra University (15%). Humanities & Arts is the area with the least output in the Web of Science, never reaching 10% of the total share of each university.

Altmetric.com covered 5922 records from the Web of Science, representing 36% of our data set (that is, 5922 publications where found to have mentions in social media and had a calculated 'altmetric' score). This share varies considerably depending on the university and research area under consideration. Figure 1 shows the share of Web of Science documents in our data set which have received at least a mention in any of the social media metrics retrieved from Altmetric.com by university. As observed, Pompeu Fabra University as the university best covered by altmetric data: 67% of all its publications received mentions in social media. The other three universities have well under 50% of their total output mentioned in social media, with values between 30-40% for Granada and Valencia, and 23% for Carlos III.

Table 2. Publication profile of four Spanish universities according to the number of papers indexed in the Web of Science in 2014

| Research Area | University of Granada | | Pompeu Fabra University | |
|---|---|---|---|---|
| | Nº WoS Documents | % WoS Documents | Nº WoS Documents | % WoS Documents |
| Engineering & Technology | 483 | 20% | 107 | 10% |
| Humanities & Arts | 84 | 4% | 59 | 6% |
| Science | 1896 | 79% | 815 | 79% |
| Social Sciences | 290 | 12% | 156 | 15% |
| **Total (without duplicates)** | **2387** | **100%** | **1029** | **100%** |

| Research Area | Polytechnic University Valencia | | Carlos III University | |
|---|---|---|---|---|
| | Nº WoS Documents | % WoS Documents | Nº WoS Documents | % WoS Documents |
| Engineering & Technology | 779 | 44% | 413 | 51% |
| Humanities & Arts | 16 | 1% | 39 | 5% |
| Science | 1272 | 73% | 365 | 45% |
| Social Sciences | 103 | 6% | 164 | 20% |
| **Total (without duplicates)** | **1753** | **100%** | **810** | **100%** |

If we consider the altmetric coverage by research area and university (figure 1, Science and Social Sciences have a similar coverage of altmetric mentions in the case of the University of Granada (34% and 31% respectively). The area of Science is the best covered for Pompeu Fabra (up to 70%), the only area where the share of mentioned papers surpasses 50%. It is followed by Social Sciences (49%). In the case of the Polytechnic University of Valencia and Carlos III University, Social Sciences is the area best covered. Humanities & Arts is again the research area less well-covered with the exception of Carlos III, where Engineering & Technology show the lowest values.

Figure 2 introduces the altmetric score as a proxy of the intensity of mentions in social media to publications. Here the prevalence of Science in Pompeu Fabra and Granada is evident. Again a pattern can be observed with regard to the areas which receive more mentions in social media: Science first, Social Sciences second, then Engineering & Technology, and Humanities & Arts. In this case, the exception can be found in Polytechnic University of Valencia, where the second area with the highest intensity of social media metrics is Engineering & Technology and not Social Sciences.

Figure 1. Coverage by fields of altmetric indicators by areas based on the number of documents published by four Spanish universities according to the Web of Science in 2014

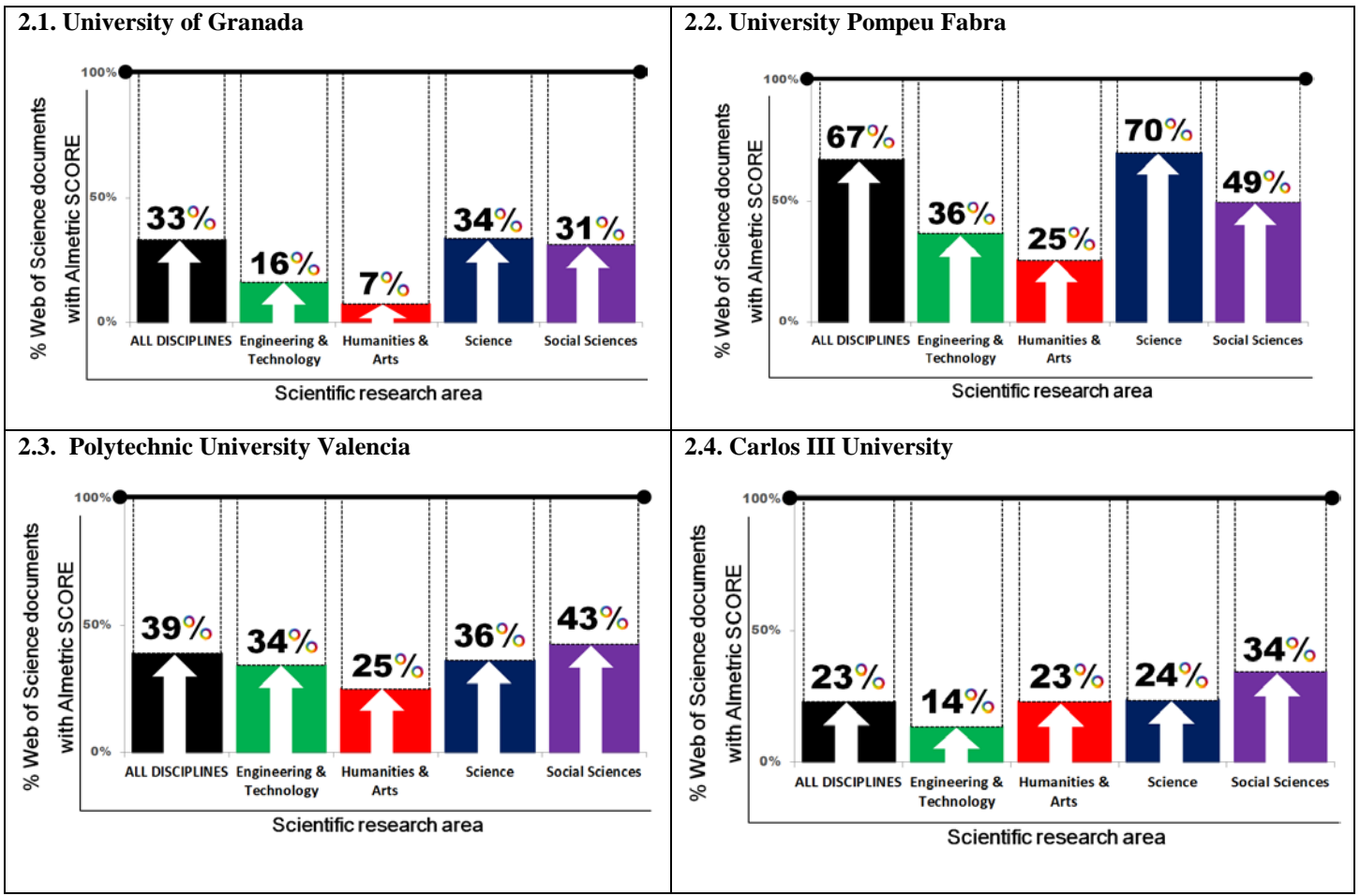

## DISCUSSION AND CONCLUDING REMARKS

This paper presents a first attempt at analysing altmetric indicators at the institutional level. Specifically, our goal was to analyze to what extent did Altmetric.com showed an adequate and homogeneous coverage among research areas. First, we observe a low coverage of altmetric indicators with only 36% of all documents retrieved from the Web of Science having an 'altmetric' score. We only find reasonable levels of coverage for Pompeu Fabra University, where 67% of all documents had altmetric mentions. This university represents a different profile to that of the other three universities analysed, confirming that this university represents an outlier of the Spanish University system (Robinson-García, Rodríguez-Sánchez, García, Torres-Salinas, & Fdez-Valdivia, 2013).

Figure 2. Boxplot with the altmetric score distribution for four Spanish universities by areas in 2014

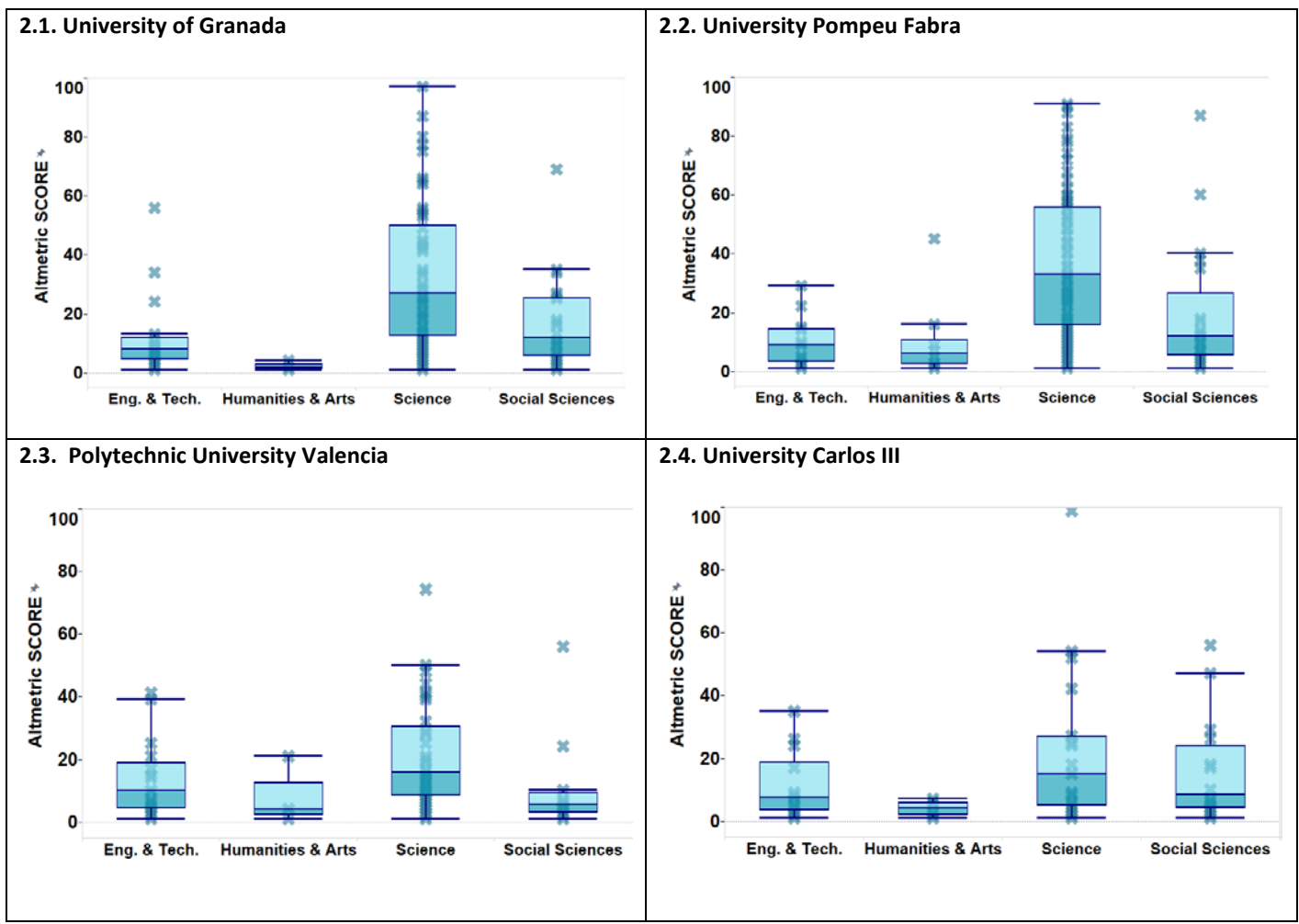

One of the main factors that may contribute to such low coverage could be the use of the DOI number as reference for querying the Altmetric.com API. One of the first problems encountered here and in any other altmetric study (Robinson-García et al., 2014)**,** is the reliance on DOI numbers to retrieve social media mentions, this assumes a necessary loss of information. First, we must note that not all papers indexed in the Web of Science include DOI (7% of the records in our data set did not include a DOI number). Second and more importantly, we must stress that Altmetric.com uses the DOI for its searches but that 1) not all social media mentions directed to a publication include the DOI number (for instance, Twitter links could use not normalized web links), and 2) not all mentions are directed to the journal article but to other versions of the same publication (i.e., post-refereed versions uploaded to a repository). This translates in a low coverage of altmetric indicators which may be misleading.

Second, we observe that for the four universities analysed, the area of Science shows higher 'altmetric' scores that the rest of the research areas. Science is also the area best covered for three of the four universities. Only in the case of the Carlos III University, Social Sciences are best covered. Despite this fact, we do not observe for any of the universities analysed a predominance of altmetric data for the areas of Social and Humanities & Arts as suggested elsewhere (Costas et al., 2015a). We could speculate that such differences between our results and previous studies could be due to the fact that none of the studied universities belongs to an English-speaking country.

However, further research is needed to gain more insight as to what is occurring. In this sense, we propose the following lines of work:

1. *The national factor of research areas.* It is necessary to verify if there is a national or linguistic factor influencing the coverage of altmetric indicators, especially in the areas of Social Sciences and Humanities & Arts. Another explanation could be that these areas are represented in social media by small communities, limiting their capacity to disseminate their research papers in social media.

2. *The use of social media by researchers.* It is important to note the influence that the dissemination of research papers by the authors themselves may play in this process. It would be interesting to analyse if universities with an active academic staff in social media could lead to a better coverage of altmetric indicators. Also the contrary should be considered. Are universities with personnel with no presence in social media worse covered by altmetric data?

3. *Collaboration networks.* Another factor that may affect coverage by areas could be international collaboration rates. Areas with more authors per paper could be better covered by social media.

Finally, considering the low coverage of altmetric data at the institutional level, it could be interesting for research policy makers to consider the development of guidelines and best practices guides to ensure that researchers disseminate adequately their research findings through social media, emphasizing the use of normalised identifiers (DOI numbers, ArXiV, PubMedID) in order to ensure the recollection of such metrics.

**Acknowledgments**
The authors would like to thank Matthew Spehlmann and Altmetric.com for the collection of the data.